\documentclass[prb,aps,twocolumn,showpacs,preprintnumbers,amsmath,amssymb,floatfix,citeautoscript]{revtex4}

\usepackage{graphicx}
\usepackage{dcolumn}
\usepackage{bm}

\begin{document}


\title{Free energy analysis of system comprising biased atomic force\\ microscope tip, water meniscus and dielectric surface}

\author{Sergei F. Lyuksyutov}
 \email{sfl@physics.uakron.edu}
\author{Pavel B. Paramonov}
	\affiliation{Departments of Physics and Polymer Engineering, The University of Akron, Akron OH 44325}
\author{Richard A. Vaia}
	\affiliation{Materials and Manufacturing Directorate, Air Force Research Laboratory, 
Wright-Patterson Air Force Base, OH 45433}

\date{\today}

\begin{abstract}
We are concerned with free energy analysis of the system comprising an AFM tip, water meniscus, and polymer film. Under applied electrostatic potential, the minimum in free energy is at a distance greater than the initial tip--substrate separation in the absence of potential. This equilibrium distance, $t_0$, mostly depends on the tip bias $V$ and cantilever spring constant $k_s$, where as variations of $t_0$ is less pronounced with respect to the dielectric constants, and polymer film thickness. Polarization of water meniscus under the AFM tip appears to be the dominant factor enabling the creation of mechanical work for tip retraction.
\end{abstract}


\maketitle

\section{Introduction}
\label{sec:intro}

The goal of this work is to analyze the free energy of the system comprising an electrically biased AFM tip, water meniscus, and dielectric surface. It is important to identify the key factors leading to the effective tip--substrate repulsion through an equilibrium free--energy balance. Complete understanding of the process is complicated by the fact that the specific spatial details of the tip--surface contact profile as well as cantilever motion {\itshape with applied bias} is exceedingly difficult to observe and quantify; and thus is not well understood. Although in this work we concentrate on the theoretical consideration of this process for polymer films in the context of the AFM-assisted electrostatic nanolithography\cite{our_AFMEN_nature}, the equilibrium free energy balance and associated conclusions are generally applicable to other material systems including semiconductors, bio-macromolecules and self--assembled monolayers.

\section{Model Description}
\label{sec:model}

\begin{figure}
\includegraphics[width=7cm]{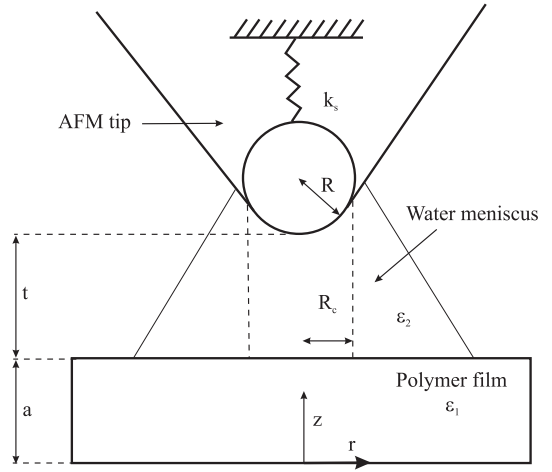}
\caption{Conceptual presentation of AFM tip--polymer surface junction.}
\label{fig:tip_lift_model_geom}
\end{figure}

Consider the schematic representation in Figure~\ref{fig:tip_lift_model_geom} as summarizing the key components involved in the process. The system comprises the AFM tip, a dielectric film, and the water meniscus filling the junction. The tip is presented as a sphere of radius $R$ attached to the cantilever with the spring constant $k_s$. The region of interest is near the tip apex. Tip-surface separation $t$ confines this region in the vertical direction, and the cylinder of the radius $R_c$ defines its lateral dimensions. The total free energy of the system can be presented as the sum of four terms:
\begin{equation}
	F = F_1 + F_2 + U + F_{ts}
\label{F_contribss}
\end{equation}
Here, $F_1$ and $F_2$ are Helmholtz's free energy contributions related to polarization of the dielectric film, and water meniscus respectively; $U$ is the energy of the contracted spring, and $F_{ts}$ is the energy of electrostatic attraction between the tip and the film. The free energy of the system is chosen to be zero for an unbiased tip at $t=0$ and thus $F$ represents the change in free energy. Free energy densities variations at the tip/water and the water/polymer interfaces due to electric field, and possible charge carriers' generation are not accounted in the Eq.~(\ref{F_contribss}). 

We consider the terms $F_1$ and $F_2$ separately, since the variation of dielectric medium entropy with respect to external electric field is greater for water than that for polymer. Additionally, we neglect the term related to Van der Waals interactions between the tip and the surface: $U_{VdW}=-AR/6t$ because the typical value of Hamaker constant for condensed matter is in the order of $10^{-19}\;J$\cite{Capella_fd_review} resulting in energy less than $10^{-29}\;J$, which is found to be negligible in comparison to the other free energy terms in (\ref{F_contribss}). 

The dielectric film is considered as a non-conductive polymer with a macroscopic dielectric constant $\epsilon_1$. The first term thus can be written as: $F_1 = \int\limits_{V}f_1dV$, where the free energy density $f_1$ is
\begin{equation}
	f_1=-\frac{\epsilon_0\epsilon_1 E^2}{2}
\end{equation}
where $E=-\nabla\phi$ is the external electric field, $\epsilon_0$ is the dielectric permittivity in free space, and we assume that $\epsilon_1$ is independent of $E$ even for the substantial fields ($\sim10^9\;Vm^{-1}$) considered. 

To account for the polarization of a liquid medium, the free energy of water has a form
\begin{equation}
	F_2=-k_B T \ln Z
\end{equation}
where $Z$ is the total partition function, $k_B$ is Boltzman's constant, and $T$ is the temperature. The partition function can be written for the canonical ensemble of $N$ molecules of liquid as $Z=Z_c\;Q^N/N!$\cite{book_statmech_Hill}. Here, $Q$ is the partition function of a molecule, and $Z_c$ is the configurational integral describing intermolecular interaction. The major contributions to the intermolecular interactions in water are hydrogen bonding, dipole-dipole and quadrupole interactions. The configurational integral $Z_c$ accounts these interactions. The intermolecular interactions in water are stronger than those between a water molecule and a field up to the field's strength of $2\cdot10^9\;Vm^{-1}$\cite{MD_water_E_Vegiri}. It is assumed that spatial arrangements of the molecules and density of intermolecular hydrogen bonds are weakly affected by electric field below $10^9\;Vm^{-1}$. Thus the variations of the configurational integral $Z_c$, and the translational contribution $q_{tr}$ to the partition function with the external field strength are neglected. Translational, internal and rotational degrees of freedom contribute to the partition function $Q$ of the molecule. The internal degrees of freedom (vibrational and electronic) are only weakly dependent on external electric field. In contrast, the molecular dipole parallel to the $C_{2V}$ symmetry axis of water molecule will affect the rotational degrees of freedom in an applied field. The above-stated assumptions allow separation of the field-independent (translational and internal) and field-dependent (rotational) contributions to the partition function:
\begin{equation}
	Q=q_{tr}\;q_{in}\;q_E
\end{equation}
where $q_{tr}$ and $q_{in}$ are the translational and internal molecule's partition functions.  The rotational function $q_E$ for the molecular dipole moment $\mu$ in the external electric field $E$ can be presented as
\begin{equation}
	q_E = \frac{\sinh(y)}{y}, \;\;\;\;\; y = \frac{\mu E}{k_B T}
\end{equation}
The free energy density change with respect to the state $E=0$ at arbitrary state of interest is found to be
\begin{equation}
	f_2 = -\frac{R_gT}{V_m}\ln q_E 
\end{equation}
where $q_{tr}$ and $q_{in}$ are assumed to be independent of the electric field. The contribution $F_2$ to the total free energy can be found by integrating the density $f_2$ over the volume: $F_2 = \int\limits_{V}f_2dV$.  

Note that the energy densities $f_1$ and $f_2$ are the functions of the coordinates $(r,z)$ inside the polymer and water meniscus because of the spatial non-uniformity of electric field created by the tip-substrate geometry. 

The free energy term associated with the contraction of the spring that represents the cantilever bending has the form
\begin{equation}
	U = \frac{k_s t^2}{2}
\end{equation}
As a first approximation, anharmonism of the spring constant $k_s$ is ignored. 

Finally, the free energy of the tip--substrate electrostatic attraction depends on the details of the electric potential (and the field) inside and outside the film. This was determined\cite{our_triboelectrification} through the exact analytical solution of the corresponding Laplace's equation using the method of images.

The energy of the tip--sample electrostatic attraction, $F_{ts}$, was calculated using the image charge construction from Ref.~[\onlinecite{our_triboelectrification}]. The result has the following form:
\begin{equation}
	F_{ts} = -\:\pi\epsilon_0 V^2 R^2 \left( \frac{\eta}{h}-\left(1-\eta^2\right) \sum_{n=0}^{\infty} \frac{\eta^n}{(n+1)a+h} \right)
\label{F_ts}
\end{equation}
Here, $V$ is the bias voltage, $a$ is the polymer film thickness, $h=t+R$, $t$ is the tip--surface distance, $R$ is the tip radius, and $\eta$ is given by
\begin{equation}
	\eta = \frac{\epsilon_2-\epsilon_1}{\epsilon_1+\epsilon_2}
\end{equation}
The dielectric constant of water, $\epsilon_2$, is, strictly speaking, a function of the electric field because dielectric saturation effects become significant at the field strength of our interest. For small field's magnitude the water's dielectric response is linear with $\epsilon=80$. The dielectric saturation effect becomes substantial as the field grows stronger than $10^7-10^8\;Vm^{-1}$ resulting in decrease of the dielectric constant with the field down to value of $2$. The molecular dynamics simulations, in particular, provide insight into dielectric saturation phenomena\cite{MD_water_eps_Sutmann}. We treat $\epsilon_2$ as a constant in the calculations of electric field distribution and free energy $F_2$. In order to investigate the real system behavior, an {\itshape effective} value of $\epsilon_2$ is varied between the low and high field limits.

We use the following dimensionless spatial variables: $\rho=r/R$, $\zeta=z/R$, and the dimensionless electric field strength $\psi(\rho,\zeta)=E R/V$. The following dimensionless parameters are defined:
\begin{equation}
	\kappa=\frac{k_sR}{\epsilon_0\epsilon_1V^2}, \;\;\;\;\; w=\frac{R_gTR^2}{\epsilon_0\epsilon_1V^2V_m}, \;\;\;\;\; m=\frac{\mu V}{Rk_BT} 
\end{equation}
After introducing $U_0=\epsilon_0\epsilon_1V^2R$ to scale energy, we finally arrive with the formula for the total dimensionless free energy function expressed in the units $U_0$:
\begin{equation}
	\frac{F(t)}{U_0} = \frac{F_1}{U_0} + \frac{F_2}{U_0} + \frac{1}{2}\kappa t^2 + \frac{F_{ts}}{U_0} 
\label{F_t_result}
\end{equation}
where,
\begin{equation}
	\frac{F_1}{U_0} = -\:\pi \int\limits_{0}^{a/R}d\zeta \int\limits_{0}^{R_c/R}\rho\;\psi^2(\rho,\zeta)d\rho
\label{F_1_result}
\end{equation}
is the free energy of the dielectric, and
\begin{equation}
	\frac{F_2}{U_0} = -\:2\pi w \int\limits_{a}^{(a+t)/R}d\zeta \int\limits_{0}^{R_c/R}\rho\; \ln\left[ \frac{\sinh\left(m\psi(\rho,\zeta)\right)}{m\psi(\rho,\zeta)} \right]d\rho
\label{F_2_result}
\end{equation}
is the free energy of water in the meniscus. Eqs.~(\ref{F_t_result}), (\ref{F_ts}) and (\ref{F_1_result})-(\ref{F_2_result}) allow calculation of the free energy as a function of bias voltage, dielectric properties, and tip--sample separation distance. 

\section{Results and Discussion}
\label{sec:results}

\begin{figure}
\includegraphics[width=7cm]{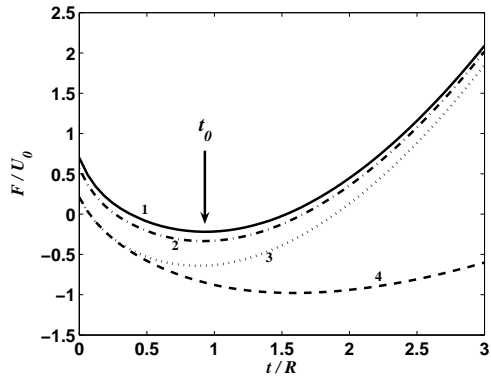}
\caption{Dependence of the dimensionless free energy function $F/U_0$ on $t/R$ for (1: solid line) bias voltage $-\;25\;V$, $\epsilon_1=3$, $\epsilon_2=80$; (2: dot--dashed line) bias voltage $-\;25\;V$, $\epsilon_1=3$, $\epsilon_2=50$; (3: dotted line) bias voltage $-\;25\;V$, $\epsilon_1=3$, $\epsilon_2=20$; (4: dashed line) bias voltage $-\;50\;V$, $\epsilon_1=3$, $\epsilon_2=20$. The following parameters used: the dipole moment of water molecule in condensed phase $\mu=1.98\;D$; the polymer film thickness $a=R=35\;nm$; the cantilever spring constant $k_s=0.35\;Nm^{-1}$; $T=25\:^0C$; $R_c=R$.}
\label{fig:tip_lift_F_t}
\end{figure}

The variations of the function $F(t)/U_0$ with respect to tip--surface separation $t$ normalized on tip radius $R$ are presented in Figure~\ref{fig:tip_lift_F_t}. The following typical (reference) values have been selected: $V=25\;V$, $\epsilon_1=3$, $\epsilon_2=50$, $a=35\;nm$, and $k_s=0.35\;Nm^{-1}$ to analyze the performance of $t_0$ in poly(methyl methacrylate) (PMMA) and polystyrene (PS) films. Starting from the initial reference point at close proximity of the tip and polymer, the function decreases as the tip retracts from the surface, passes through a minimum when tip--sample separation is in the order of $30\;nm$, and then increases again. The equilibrium tip--surface separation, $t_0$, corresponds to the minimum of the function $F(t)/U_0$. Non-zero value of $t_0$ implies spontaneous tip retraction from the surface, associated with the minimization of the free energy. 

The value of the equilibrium tip--surface separation is found to be $t_0/R=0.92$. Selecting $R=35\;nm$, one could estimate $U_0=5.8\cdot10^{-16}\;J$ ($3650\;eV$). The major contributions to the system's free energy are due to the free energy of water meniscus $F_2$ ($-\:0.5\cdot10^{-15}\;J$) and the spring contraction energy $U$ ($0.18\cdot10^{-15}\;J$) - each playing against the other. The contribution of the electrostatic term $F_{ts}$ ($0.11\cdot10^{-15}\;J$) is comparable to the spring energy, whereas the term associated with polarization of polymer dielectric $F_1$ is small ($-\:0.002\cdot10^{-15}\;J$), and does not play a significant role. Thus, the shift in the equilibrium tip--surface separation, $t_0$, arises from the maximization of the volume of water within the highest field region under the tip apex -- the initial water meniscus lowers its free energy by expanding vertically.

\begin{figure}
	\begin{minipage}{7cm}
		\begin{center}
		\includegraphics[width=7cm]{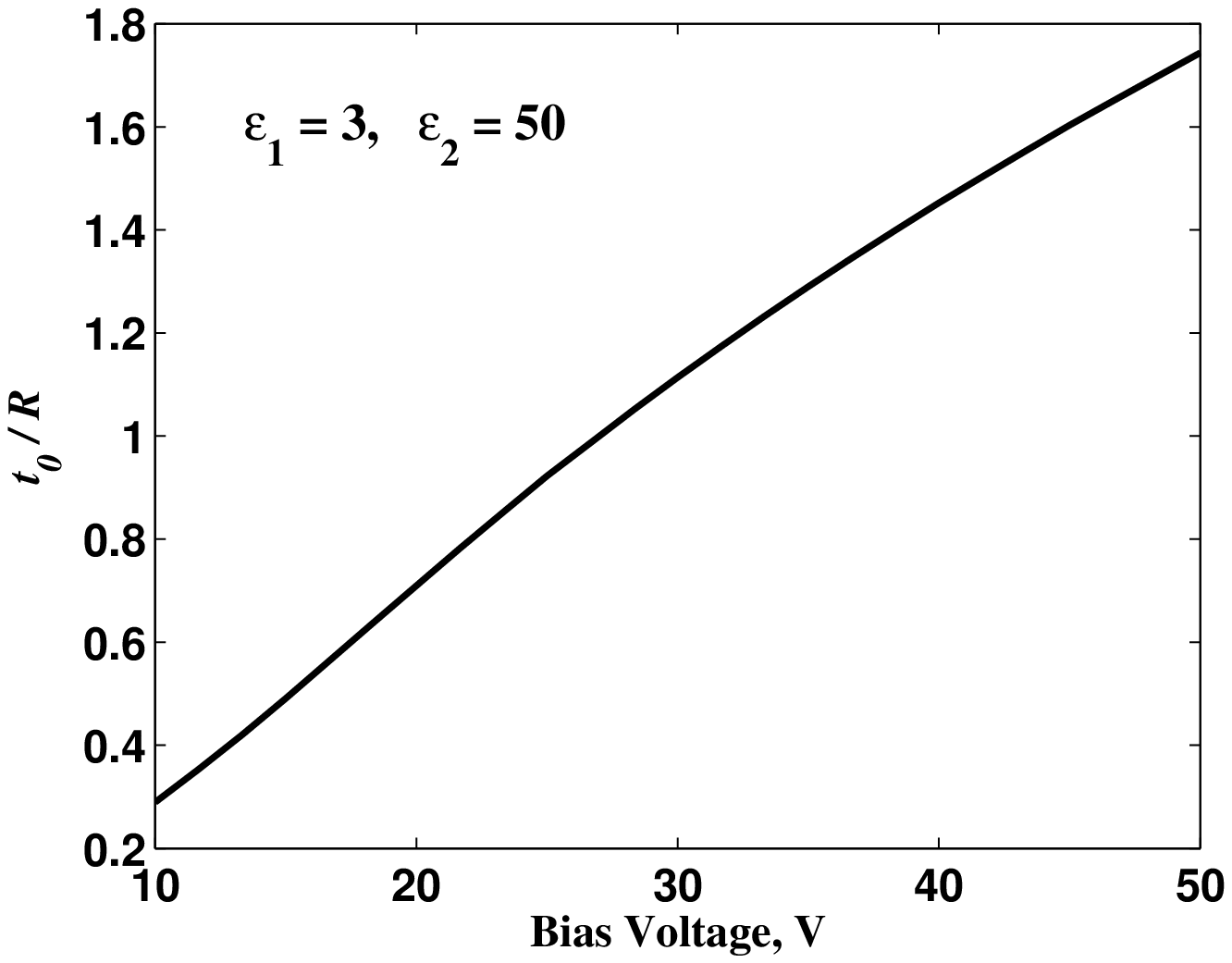}
		\end{center}
	\end{minipage} 
	\begin{minipage}{7cm}
		\begin{center}
		\includegraphics[width=7cm]{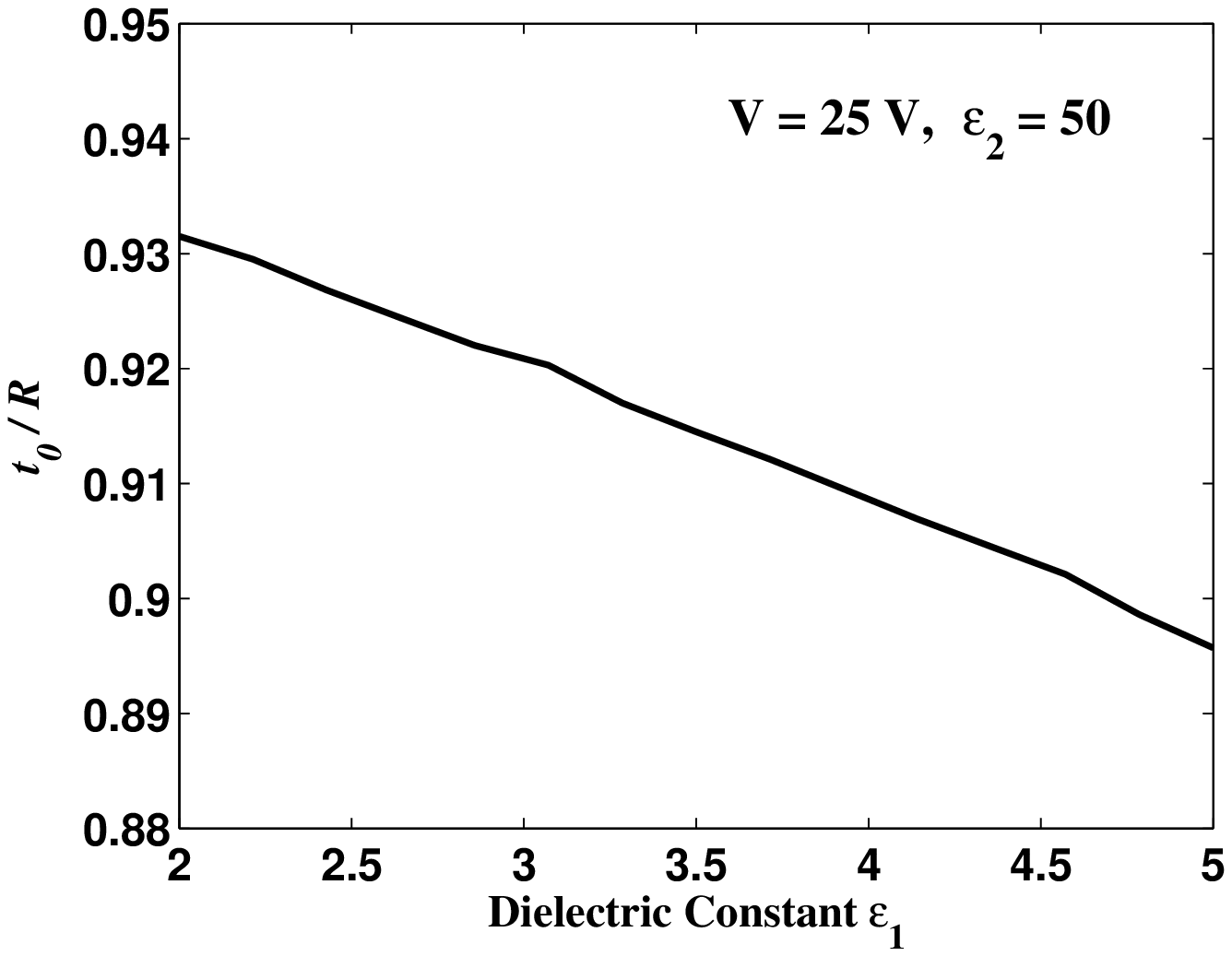}
		\end{center}
	\end{minipage} 
\caption{Variation of $t_0$ with respect to the bias voltage (for $\epsilon_1=3$, $\epsilon_2=50$) and with respect to dielectric constant of polymer film $\epsilon_1$ (for bias voltage $-\;25\;V$, and $\epsilon_2=50$).}
\label{figs:t0_V_eps1}
\end{figure}

The value of $t_0$ depends on parameters such as bias voltage $V$ and dielectric constant of polymer $\epsilon_1$ (Figure~\ref{figs:t0_V_eps1}), effective dielectric constant of water $\epsilon_2$, polymer film thickness $a$ and cantilever spring constant $k_s$. An analysis of $t_0$ has been conducted using dimensionless parameters, proportional to the first derivative of $t_0$ with respect to the corresponding quantities, calculated in the vicinity of the reference values mentioned above. This analysis indicates that $t_0$ mostly depends on the tip bias $V$ and cantilever spring constant $k_s$. The variations of $t_0$ with respect to the dielectric constants, and polymer film thickness found to be less pronounced. Therefore, the nanostructures are expected to be tunable simply adjusting the bias voltage, the easiest parameter to control experimentally. This trend seems to be general for different parameter's values used in this model.

\section{Conclusions}

The analysis of the free energy of the system comprising an electrically biased AFM tip, water meniscus, and polymer film indicates that the equilibrium tip--surface distance is comparable to the tip's radius. The mechanical work to lift the tip is produced by the volume of water penetrating the tip--surface junction. Another reason is associated with the tip repulsion from the surface in the double--layered (water and polymer) system. The sign of the force acting on the charge $\cal Q$ immersed inside the dielectric with permittivity $\epsilon_2$ at a certain distance from the dielectric with another permittivity $\epsilon_1$ depends on the sign of $\epsilon_2-\epsilon_1$: repulsive for the positive sign, and attractive for the negative one. Thus the peculiar behavior of the AFM tip can be attributed to energy transfer from electrostatic field increasing the potential energy of AFM cantilever. This effect presents an additional opportunity for high aspect ratio nanostructures formation.

\begin{acknowledgments}
Authors thank Dr. Grigori Sigalov for discussions.
\end{acknowledgments}

\bibliography{bio}

\end{document}